\documentclass[11pt,aps,prd,onecolumn,tightenlines,superscriptaddress,amssymb,amsmath,nofootinbib,showkeys]{revtex4-2}

\usepackage[T1]{fontenc}
\usepackage{lmodern}
\usepackage[letterpaper,hmargin=1.25in,vmargin=1.1in]{geometry}
\usepackage{graphicx}
\usepackage{bm}
\usepackage{xcolor}
\usepackage{microtype}
\usepackage{hyperref}
\usepackage{orcidlink}

\definecolor{linkblue}{RGB}{0,51,153}
\hypersetup{
    colorlinks=true,
    linkcolor=linkblue,
    filecolor=linkblue,
    citecolor=linkblue,
    urlcolor=linkblue,
    pdftitle={A friendly phantom: Late-time AdS-to-dS transition and cosmological tensions},
    pdfauthor={Özgür Akarsu; Leandros Perivolaropoulos; A. Emrah Yükselci; Alexander Zhuk}
}

\begin{document}
\linespread{1.06}\selectfont

\title{A friendly phantom: Late-time AdS-to-dS transition\\ and cosmological tensions}

\author{\"{O}zg\"{u}r Akarsu\,\orcidlink{0000-0001-6917-6176}}
\email{akarsuo@itu.edu.tr}
\affiliation{Department of Physics, Istanbul Technical University, Maslak 34469 Istanbul, Turkey}

\author{Leandros Perivolaropoulos\,\orcidlink{0000-0001-9330-2371}}
\email{leandros@uoi.gr}
\affiliation{Department of Physics, University of Ioannina, GR-45110 Ioannina, Greece}

\author{A. Emrah Y\"{u}kselci\,\orcidlink{0000-0003-4554-1268}}
\email{yukselcia@itu.edu.tr}
\affiliation{Department of Physics, Istanbul Technical University, Maslak 34469 Istanbul, Turkey}

\author{Alexander Zhuk\,\orcidlink{0000-0001-9107-5048}}
\email{ai.zhuk2@gmail.com}
\thanks{Corresponding author.}
\affiliation{Deutsches Zentrum f\"ur Astrophysik, Postplatz 1, 02826 G\"orlitz, Germany}
\affiliation{Astronomical Observatory, Odessa National University, Dvoryanskaya Street 2, 65082 Odessa, Ukraine}

\begin{abstract}
We present Ph-$\Lambda_{\rm s}$CDM, a phantom-scalar realization within general relativity of the sign-switching cosmological-constant idea, $\Lambda_{\rm s}$CDM, in which a phantom scalar evolving on a bounded hyperbolic-tangent potential induces a smooth mirror AdS-to-dS transition in the late-time dark-energy density. The wrong-sign kinetic term, usually viewed as pathological, becomes the mechanism lifting the field from a negative- to a positive-energy vacuum-like regime. The construction also shows that the field can become repulsive while its energy density is still negative. The cosmology nevertheless remains controlled: total energy stays positive, the late-time attractor is de Sitter rather than a Big Rip, and the dynamics remain safely infrared. Ph-$\Lambda_{\rm s}$CDM thus offers a concrete late-time mechanism with the potential to address multiple cosmological tensions.
\end{abstract}


 \maketitle
\renewcommand{\thefootnote}{\alph{footnote}}

\begin{center}
\begin{minipage}{0.7\linewidth}
\centering\small\emph{This essay received an Honorable Mention in the 2026 Essay Competition of the Gravity Research Foundation.}
\end{minipage}
\end{center}
\vspace{0.75em}

In modern cosmology, phantom matter typically enters the discussion as a warning sign---a \emph{menace}, in Caldwell's phrase---rather than as a candidate solution. Since Caldwell's original discussion of components with equation of state (EoS) $\omega<-1$, wrong-sign kinetic terms have been associated with violations of the energy conditions, Big Rip singularity, tachyonic instability and vacuum instability at the fundamental level~\cite{Caldwell:1999ew,Caldwell:2003vq,Carroll:2003st,Cline:2003gs,Arkani-Hamed:2003pdi,Piazza:2004df}. Yet the same sign reversal gives a phantom field one remarkable property unavailable to canonical scalars: in an expanding universe it can climb, rather than roll down, a potential.

This observation becomes especially interesting in the context of the sign-switching cosmological-constant framework $\Lambda_{\rm s}$CDM, among the most economical late-time extensions of the standard $\Lambda$ cold dark matter ($\Lambda$CDM) model, proposed so far to address the cosmological tensions that have persistently emerged over the past decade in the era of precision cosmology~\cite{Perivolaropoulos:2021jda,Abdalla:2022yfr,Akarsu:2024qiq,CosmoVerseNetwork:2025alb}. In its original phenomenological form, $\Lambda_{\rm s}$CDM posits a rapid mirror AdS-to-dS transition around $z_{\dagger}\sim2$, with the effective cosmological constant changing sign from negative to positive while preserving its characteristic magnitude~\cite{Akarsu:2019hmw,Akarsu:2021fol,Akarsu:2022typ,Akarsu:2023mfb}. A growing body of analyses indicates that this framework can alleviate multiple cosmological tensions, including those in $H_0$, the closely related $M_{\rm B}$, $S_8$ and the growth index $\gamma$~\cite{Akarsu:2019hmw,Akarsu:2021fol,Akarsu:2022typ,Akarsu:2023mfb,Paraskevas:2024ytz,Akarsu:2025ijk,Escamilla:2025imi,Akarsu:2025nns}. It is also of particular interest because it can yield an age of the universe consistent with estimates from the oldest globular clusters~\cite{Akarsu:2022typ} and, when neutrino properties are allowed to vary, do so without requiring departures from the standard expectations for $N_{\rm eff}$ and $\sum m_\nu$~\cite{Yadav:2024duq}, thereby suggesting that it may also help address the recently emerged effective negative-neutrino-mass anomaly~\cite{Elbers:2024sha,Elbers:2025vlz}.

The physical logic is transparent. If dark energy (DE) had negative energy density before the transition, then the expansion rate is suppressed relative to $\Lambda$CDM at $z>z_{\dagger}$; consistency with the CMB distance to last scattering then demands compensating enhancement at lower redshifts, which naturally pushes the inferred $H_0$ upward~\cite{Akarsu:2019hmw,Akarsu:2021fol,Akarsu:2022typ}. In its abrupt signum form, the $\Lambda_{\rm s}$CDM model, $\Lambda_{\rm s}(z)=\Lambda\,{\rm sgn}(z_{\dagger}-z)$, is only a phenomenological idealization of a rapid transition, but it makes the mirror character of the transition explicit and provides the limiting case that the smooth phantom realization seeks to reproduce~\cite{Akarsu:2019hmw,Akarsu:2021fol,Akarsu:2022typ,Akarsu:2023mfb}. The central theoretical question is therefore whether such a background history can be realized within general relativity (GR), i.e. without modifying the Einstein gravitational sector. While $\Lambda_{\rm s}$CDM was originally conjectured phenomenologically from the findings of graduated dark energy (gDE)~\cite{Akarsu:2019hmw}, several theoretical realizations of the framework have since been proposed, including the string-inspired $\Lambda_{\rm s}$CDM$^+$~\cite{Anchordoqui:2023woo,Anchordoqui:2024gfa,Anchordoqui:2024dqc,Soriano:2025gxd}, the type-II minimally modified gravity realization $\Lambda_{\rm s}$VCDM~\cite{Akarsu:2024qsi,Akarsu:2024eoo,DeFelice:2020eju}, the teleparallel realization $f(T)$-$\Lambda_{\rm s}$CDM~\cite{Akarsu:2024nas,Souza:2024qwd} and, in certain formulations of GR, a construction based on an overall sign change of the metric~\cite{Alexandre:2023nmh}. In this essay, we focus on the recently proposed Ph-$\Lambda_{\rm s}$CDM model, which provides a smooth realization of this scenario within GR by means of a phantom scalar field evolving on a suitable potential~\cite{Akarsu:2025gwi,Akarsu:2025dmj,Adil:2026kfn}.

For canonical scalar fields, the mirror AdS-to-dS transition is dynamically disfavored: the field is driven toward lower, not higher, potential energy. A phantom field reverses this logic. Because the sign of its kinetic term flips the sign of the effective force in the Klein--Gordon equation, the field can climb a suitably chosen potential and thereby lift the DE density from negative to positive values. In this sense, the usual objection to phantom matter becomes the very mechanism required by the $\Lambda_{\rm s}$CDM framework~\cite{Akarsu:2025gwi}. Following Ref.~\cite{Akarsu:2025gwi}, we refer to this GR phantom-scalar realization of the $\Lambda_{\rm s}$CDM framework as Ph-$\Lambda_{\rm s}$CDM.

We consider a model in which the gravitational sector is described by the standard Einstein--Hilbert action of GR, while the usual positive (de Sitter-like) cosmological constant of $\Lambda$CDM is replaced by a minimally coupled phantom scalar field; the total action reads
\begin{equation}
\label{eq:action}
S=\frac{1}{c}\int {\rm d}^4x\,\sqrt{-g}\left[\frac{R}{2\kappa}+\mathcal{L}_{\phi}\right]+S_{\rm m}+S_{\rm r}\, ,
\end{equation}
where $\kappa \equiv 8\pi G_N/c^4$, with $G_N$ the Newton constant ($\kappa=M_{\rm Pl}^{-2}$ in units $\hbar=c=1$, where $M_{\rm Pl}$ is the reduced Planck mass), and $S_{\rm m}$ and $S_{\rm r}$ are the standard actions of pressureless matter and radiation. Thus, by ``within GR'' we mean that the gravitational sector is not modified; we do not mean that GR alone generates the transition. The new ingredient is an additional, minimally coupled---and admittedly exotic---matter degree of freedom. With metric signature $(+---)$, the phantom scalar field has Lagrangian density
\begin{equation}
\label{eq:lagrangian}
\mathcal{L}_{\phi}= -\frac{1}{2}g^{ik}\partial_{i}\phi\partial_{k}\phi -V(\phi)\, ,
\end{equation}
where the potential takes the hyperbolic-tangent form
\begin{equation}
\label{eq:potential}
V(\phi)=\Lambda\left[\frac{\alpha}{2} +\left(1-\frac{\alpha}{2}\right)\tanh (\nu\phi/M_{\rm Pl})\right] \, ,\quad \Lambda>0\, .
\end{equation}

The hyperbolic-tangent potential is a phenomenological yet analytically clean proof-of-concept for a smooth mirror AdS-to-dS transition in GR. In the mirror case $\alpha=0$, it interpolates between a negative plateau and a positive plateau, allowing the field to begin in an AdS-like phase and settle at late times on a dS-like plateau. The boundedness of the potential is crucial: it prevents unbounded growth of the phantom energy and makes a controlled late-time de Sitter attractor possible. More generally, $\alpha$ labels a broader transition family, with $\alpha=0$ giving the mirror case studied here, $\alpha=1$ an emergent cosmological-constant case, and $\alpha=2$ the $\Lambda$CDM limit~\cite{Akarsu:2025gwi}. We emphasize that no fixed parameter or coupling changes sign in this construction: $\Lambda$ in Eq.~\eqref{eq:potential} is a fixed positive constant, and it is the emergent DE density $\varepsilon_\phi$ that switches sign dynamically as the field climbs the potential. The sign switch is therefore not inserted as a time-dependent fundamental cosmological constant. In contrast to other realizations of the $\Lambda_{\rm s}$CDM framework, which modify the gravitational sector or the spacetime structure~\cite{Akarsu:2024qsi,Akarsu:2024eoo,DeFelice:2020eju,Akarsu:2024nas,Souza:2024qwd,Alexandre:2023nmh} or rely on string-inspired ingredients~\cite{Anchordoqui:2023woo,Anchordoqui:2024gfa,Anchordoqui:2024dqc,Soriano:2025gxd}, here the mechanism operates through classical matter dynamics in Einstein gravity. Its physical price is the wrong-sign kinetic term, so the scalar is understood as a low-energy effective description whose limitations are discussed below.

At the background level, the phantom sector admits an effective perfect-fluid description with isotropic energy density and pressure
\begin{equation}
\label{eq:rho-pressure}
\varepsilon_{\phi}=-X+V(\phi), \qquad
p_{\phi}=-X-V(\phi)\, ,
\end{equation}
where $X\equiv (\dot\phi)^2/(2c^2)>0$ and the dot denotes differentiation with respect to cosmic time $t$. The associated EoS parameter is
\begin{equation}
\label{eq:eos}
\omega_\phi=\frac{p_\phi}{\varepsilon_\phi}=\frac{-X-V(\phi)}{-X+V(\phi)}\, .
\end{equation}
Crucially, because the Lagrangian density is linear in $X$, the propagation sector remains regular and the microphysical sound speed is exactly luminal: $c_s^2
=\frac{\partial p_{\phi}}{\partial \varepsilon_{\phi}}\,c^2
=\frac{p_{\phi,X}}{\varepsilon_{\phi,X}}\,c^2
=c^2$, where a comma followed by $X$ denotes differentiation with respect to $X$.

The mirror AdS-to-dS transition, $\alpha=0$, induces a field evolution from an AdS-like negative-energy regime at early times to a dS-like positive-energy one at late times. A conceptual clarification is therefore essential. The field remains phantom throughout, including during the transition, because the sign of its kinetic term never changes; equivalently, $\varepsilon_\phi+p_\phi=-2X<0$ at all times. In the usual positive-density setting, the phantom-divide line (PDL), $\omega_\phi=-1$, i.e. the EoS value associated with a cosmological constant, coincides with the null-energy-condition boundary (NECB), $\varepsilon_\phi+p_\phi=0$. Once $\varepsilon_\phi$ is allowed to change sign, however, this coincidence is lost: the same phantom field occupies the $\omega_\phi>-1$ branch when $\varepsilon_\phi<0$ and the $\omega_\phi<-1$ branch when $\varepsilon_\phi>0$, while remaining on the $\varepsilon_\phi+p_\phi<0$ side of the NECB throughout. Thus, in the present model the PDL is not a global classifier of phantom behavior, whereas the NECB is (see Refs.~\citealp{Akarsu:2026anp} and~\citealp{Gokcen:2026pkq}). No actual NECB crossing occurs anywhere along the evolution.

It is also important to distinguish two characteristic redshifts. In the mirror case, $z_{\rm t}$ is defined by $V(z_{\rm t})=0$, whereas $z_{\dagger}$ marks the moment at which $\varepsilon_\phi(z_{\dagger})=0$. Since $\varepsilon_\phi=-X+V$ with $X>0$, the energy-density crossing occurs only after the potential has already become positive, hence $z_{\dagger}<z_{\rm t}$. More generally, the transition in the energy density is broader than the transition in the potential itself: the expansion history responds to $\varepsilon_\phi=-X+V$, not to $V$ alone. The origin of both features is made explicit by the Klein--Gordon dynamics.

The Klein--Gordon equation for the scalar field is
\begin{equation}
\label{eq:kg}
\ddot\phi+3H\dot\phi - c^2\frac{{\rm d}V}{{\rm d}\phi}=0\, ,
\end{equation}
where $H=\dot a/a$ is the Hubble parameter. Equation~\eqref{eq:kg} makes explicit not only why the phantom realization works but also why the transition in $\varepsilon_\phi$ is broader than that in $V$. For a canonical scalar, the force term enters with the opposite sign, so the field is driven down the potential; here the reversed sign makes the field climb it instead, enabling the mirror AdS-to-dS transition. As the field enters the steep part of the potential, the reversed force accelerates it uphill, increasing $X$ and partly offsetting the growth of $V$; the sign change of $\varepsilon_\phi$ is therefore postponed relative to that of the potential. After the transition, once the field reaches the nearly flat positive plateau, the force term is suppressed, but the kinetic contribution does not vanish abruptly; instead, Hubble friction damps $\dot\phi$ gradually. Over the transition interval, the residual contribution $-X$ therefore persists after most of the change in $V$ has occurred. As a result, the response of $\varepsilon_\phi$, and therefore of $H(z)$, is spread over a wider redshift interval than the potential step itself, while the evolution is steered toward a dS-like attractor rather than a Big Rip.

We supplement the phantom sector with pressureless matter (CDM+baryons) and radiation, with $p_{\rm m}=0$, $p_{\rm r}=\varepsilon_{\rm r}/3$, and energy densities $\varepsilon_{\rm m}=\varepsilon_{\rm m0}/a^3$ and $\varepsilon_{\rm r}=\varepsilon_{\rm r0}/a^4$, adopting the normalization $a_0=1$ today. The Friedmann and Raychaudhuri equations then read
\begin{equation}
\label{eq:friedmann-raychaudhuri}
\frac{3H^2}{c^2} = \kappa \left(\varepsilon_{\rm m}+\varepsilon_{\rm r}+\varepsilon_\phi\right), \qquad
\frac{\ddot a}{a} =
-\frac{\kappa c^2}{6}
\left[
(\varepsilon_{\rm m}+3p_{\rm m})
+(\varepsilon_{\rm r}+3p_{\rm r})
+(\varepsilon_{\phi}+3p_{\phi})
\right].
\end{equation}

To connect this late-time construction to observations, we impose a conservative requirement: the pre-recombination era should remain effectively consistent with $\Lambda$CDM. Then the sound horizon at last scattering is essentially unchanged, while the high-precision CMB determination of the angular acoustic scale translates into an equally precise requirement on the comoving angular diameter distance to last scattering. At the same time, the CMB fixes the physical matter density. Accordingly, from Planck-$\Lambda$CDM~\cite{Planck:2018vyg} constraints, we impose  $D_M(z_*)=\frac{1}{H_0}\int_0^{z_*}\frac{c}{\tilde h}\,\mathrm{d}z = 13869.57\ {\rm Mpc}$ and  $\Omega_{\rm m0}h^2 = 0.14314$, where $\tilde h\equiv H/H_0$, $z_*\sim1090$, and $h\equiv H_0/(100~{\rm km\,s^{-1}\,Mpc^{-1}})$. We then ask whether a late-time sign switch can satisfy these early-universe anchors while reproducing the local SH0ES determination~\cite{Riess:2021jrx}, $H_0=73.04~{\rm km\,s^{-1}\,Mpc^{-1}}$. Once the field reaches the nearly flat positive plateau, Eq.~\eqref{eq:kg} reduces approximately to $\ddot\phi+3H\dot\phi\simeq0$, so Hubble friction rapidly damps the motion and the scalar is nearly frozen today. Consequently, for the SH0ES value one finds $\Omega_{\rm m0}\equiv\varepsilon_{\rm m0}/\varepsilon_{\rm cr0}\approx0.27$, and hence $\Omega_{\Lambda0}\equiv\Lambda/\varepsilon_{\rm cr0}\approx0.73$.

\begin{figure}[!t]
    \centering
    \includegraphics[width=0.90\linewidth]{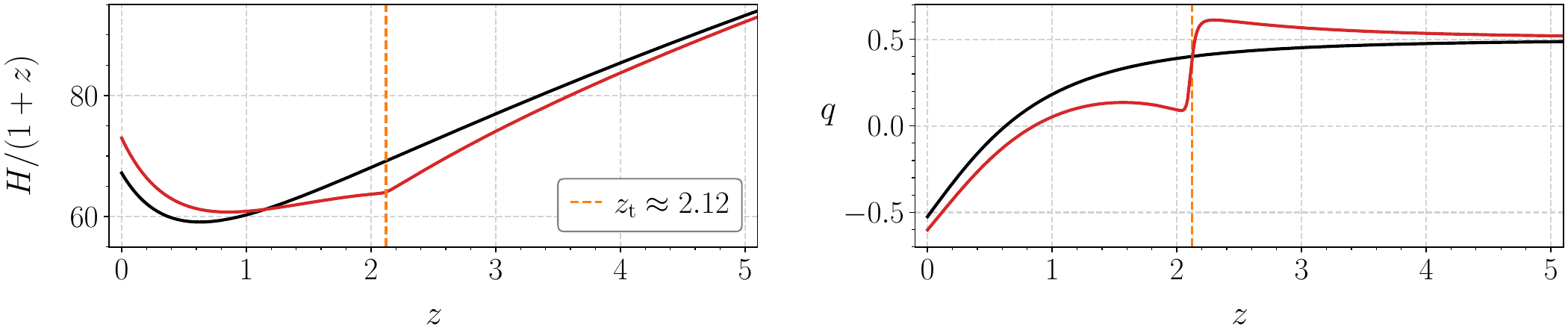}
    \caption{(Color online) Left panel: the comoving Hubble parameter $H/(1+z)=\dot a$, in units of ${\rm km\,s^{-1}\,Mpc^{-1}}$, as a function of redshift $z$. Right panel: the corresponding deceleration parameter $q(z)$. The red curves show the mirror-transition phantom solution with $\alpha=0$ and $\nu=100$, normalized to the SH0ES value $H_0=73.04~{\rm km\,s^{-1}\,Mpc^{-1}}$. The black curves show the standard $\Lambda$CDM model with $H_0=67.22~{\rm km\,s^{-1}\,Mpc^{-1}}$. The vertical dashed line marks the transition redshift $z_{\rm t}\approx2.12$, at which the potential crosses zero.}
    \label{fig:hubbles_v_100}
\end{figure}

\begin{figure}[!t]
    \centering
    \includegraphics[width=0.49\linewidth]{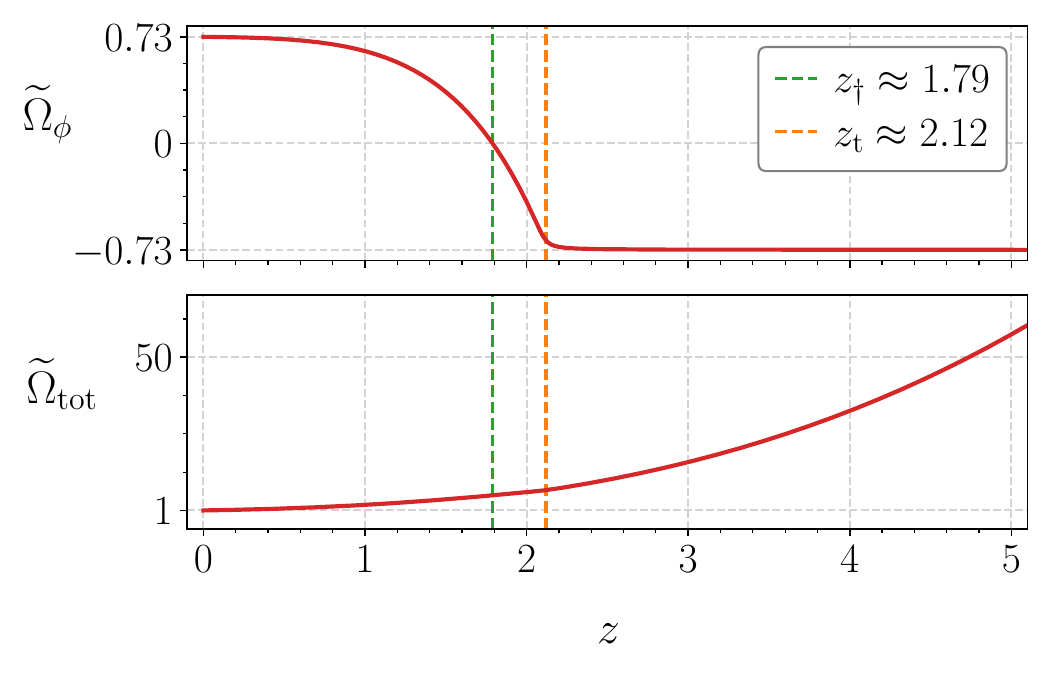}\hfill
    \includegraphics[width=0.49\linewidth]{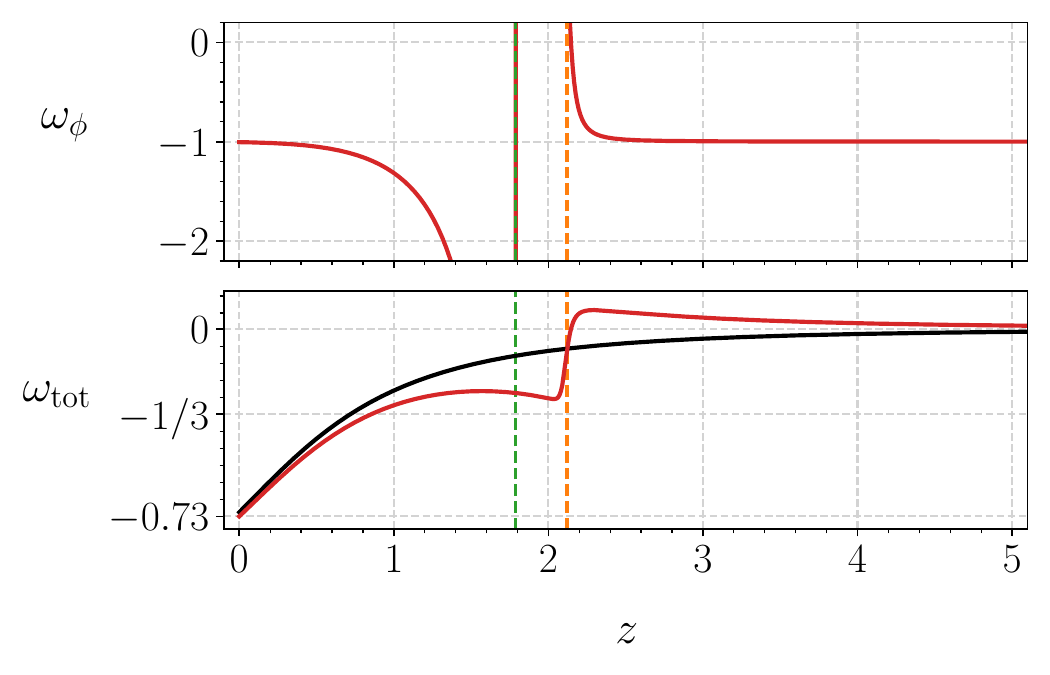}
   \caption{(Color online) Mirror-transition benchmark with $\alpha=0$ and $\nu=100$. Left composite panel: upper curve, the dimensionless phantom-sector energy density $\widetilde\Omega_{\phi}=\varepsilon_{\phi}/\varepsilon_{\rm cr0}$; lower curve, the dimensionless total energy density $\widetilde\Omega_{\rm tot}=\varepsilon_{\rm tot}/\varepsilon_{\rm cr0}$. Right composite panel: upper curve, the phantom-field EoS parameter $\omega_{\phi}$; lower curve, the effective total EoS parameter $\omega_{\rm tot}$ of the full cosmic medium (phantom + CDM + baryons + radiation). The green dashed line marks $z_{\dagger}\approx1.79$, where the phantom energy density changes sign, while the orange dashed line marks $z_{\rm t}\approx2.12$, where the potential crosses zero. The total energy density remains positive throughout the redshift interval shown.}
    \label{fig:omegas_Omegas_1}
\end{figure}

For the fiducial mirror-transition solution with $\alpha=0$ and $\nu=100$, chosen to satisfy the above Planck and SH0ES anchors, the numerical evolution yields $z_{\rm t}\approx2.12$ and $z_{\dagger}\approx1.79$, as shown in Figs.~\ref{fig:hubbles_v_100} and \ref{fig:omegas_Omegas_1}. Figure~\ref{fig:hubbles_v_100} shows the characteristic background evolution for this mirror-transition. The transition is smooth but leaves a clear dynamical imprint. For $z>z_{\dagger}$ the phantom energy density is negative, and the Friedmann part of Eq.~\eqref{eq:friedmann-raychaudhuri} therefore implies $H(z)<H(z)_{\Lambda{\rm CDM}}$. If the comoving distance $D_M(z_*)$ is to remain fixed at the CMB value, this suppression at intermediate redshifts must be compensated by a later enhancement of $H(z)$ at lower redshifts. This is the basic mechanism by which the model raises the present-day Hubble parameter while remaining consistent with the early-universe distance anchor.

Equations~\eqref{eq:rho-pressure} and \eqref{eq:eos}, together with the upper panels of Fig.~\ref{fig:omegas_Omegas_1}, make this branch structure explicit. Since $\omega_\phi+1=\frac{\varepsilon_\phi+p_\phi}{\varepsilon_\phi}
=-\frac{2X}{\varepsilon_\phi}$, with $X>0$, the sign of $\omega_\phi+1$ is controlled entirely by the sign of $\varepsilon_\phi$. Thus $\omega_\phi>-1$ on the negative-density branch $\varepsilon_\phi<0$, whereas $\omega_\phi<-1$ on the positive-density branch $\varepsilon_\phi>0$. The upper-left panel of Fig.~\ref{fig:omegas_Omegas_1} shows that the phantom energy density $\widetilde\Omega_\phi=\varepsilon_\phi/\varepsilon_{\rm cr0}$ changes sign at $z=z_{\dagger}$, while the upper-right panel shows that $\omega_\phi$ develops a pole there. This divergence is purely kinematic, not dynamical: it arises because $\omega_\phi$ is defined as the ratio $p_\phi/\varepsilon_\phi$, whereas the quantities governing the dynamics remain finite. At the crossing one has $V=X$, so that $p_\phi=-2X<0$ and $\varepsilon_\phi+p_\phi=-2X<0$, while the microphysical sound speed remains exactly $c_s^2=c^2$, as shown above. Hence the field passes smoothly through $\varepsilon_\phi=0$ even though $\omega_\phi$ itself becomes singular there.

These same panels reveal an equally important corollary concerning repulsive behavior. Between $z_{\dagger}$ and $z_{\rm t}$, the upper-left panel shows that $\varepsilon_\phi<0$, while the upper-right panel shows that $\omega_\phi$ rises above $-1/3$ over part of this interval. Therefore $\varepsilon_\phi+3p_\phi=\varepsilon_\phi(1+3\omega_\phi)<0$, so the scalar is already repulsive, contributing positively to the acceleration Eq.~\eqref{eq:friedmann-raychaudhuri}, before its energy density becomes positive. On the positive-density branch this reduces to the familiar proxy $\omega_\phi<-1/3$, but on the negative-density branch the proxy reverses and repulsion corresponds to $\omega_\phi>-1/3$. The transition therefore makes explicit that repulsive DE need not have positive energy density or satisfy the textbook inequality $\omega_\phi<-1/3$.

The lower panels of Fig.~\ref{fig:omegas_Omegas_1} also show why the dynamical behavior of the model does not render the full cosmology pathological. Although the phantom sector by itself can violate the familiar energy inequalities, the total cosmic medium composed of the phantom field, dust (CDM and baryons), and radiation remains well behaved throughout the evolution shown, from the early universe to the present: the lower-left panel shows that the total energy density $\widetilde\Omega_{\rm tot}=\varepsilon_{\rm tot}/\varepsilon_{\rm cr0}$ stays positive, while the lower-right panel shows that the effective total EoS parameter $\omega_{\rm tot}=p_{\rm tot}/\varepsilon_{\rm tot}$ remains finite and greater than $-1$. Hence $\varepsilon_{\rm tot}>0$ and $\varepsilon_{\rm tot}+p_{\rm tot}>0$, so the weak energy condition is satisfied by the total cosmic medium over this interval. Moreover, once the transition is completed, the DE sector rapidly approaches a dS-like state, its subsequent evolution becomes extremely slow, and the model avoids a Big Rip. The mechanism can be seen directly from the scalar continuity equation: $\dot\varepsilon_\phi=-3H(\varepsilon_\phi+p_\phi)=6HX\geq0$ with
$\varepsilon_\phi=-X+V\leq V\leq\Lambda$. Thus, on the expanding branch, the phantom density increases monotonically but is bounded above; it cannot undergo the runaway growth responsible for a Big Rip. In the future, matter and radiation dilute, while $\varepsilon_\phi\leq\Lambda$; hence the total energy density, and therefore $H$ through the Friedmann equation, remain bounded, so the scale factor cannot diverge at a finite cosmic time. Because the potential is smooth with bounded derivatives and the background approaches a regular de Sitter limit, the other finite-time future singularities of Types II--IV in the standard classification~\cite{Nojiri:2005sx} are also absent for the solution considered here.

The two inequalities entering the WEC for the total medium have different late-time behavior. Along the regular expanding solution, $\varepsilon_{\rm tot}=3H^2/(\kappa c^2)>0$ throughout and $\varepsilon_{\rm tot}\to\Lambda$. The second inequality, $\varepsilon_{\rm tot}+p_{\rm tot}\geq0$, holds over the history displayed in Figs.~\ref{fig:hubbles_v_100} and \ref{fig:omegas_Omegas_1}, through the present, but not indefinitely. On the exact tanh tail, $V_{,\phi}$ is exponentially small but nonzero at every finite $\phi$; the late-time slow-climb balance gives $\dot\phi\propto t^{-1}$ and hence $X\propto t^{-2}$, whereas matter and radiation dilute exponentially as the spacetime approaches de Sitter. Once $\varepsilon_{\rm m}+4\varepsilon_{\rm r}/3<2X$, one has $\varepsilon_{\rm tot}+p_{\rm tot}
=\varepsilon_{\rm m}+\frac{4}{3}\varepsilon_{\rm r}-2X<0$, and $\dot H\to0^+$. Both the phantom-sector NEC violation, $\varepsilon_\phi+p_\phi=-2X$, and the ensuing total-medium NEC violation vanish asymptotically, while $H$ remains bounded and approaches $H_{\rm dS}$ from below, without a Big Rip. In this sense, the phantom remains exotic at the component level but controlled at the level of the full FRW background. The same conclusion is reflected in Fig.~\ref{fig:hubbles_v_100}, where both the comoving Hubble parameter $\dot a=H/(1+z)$ and the deceleration parameter $q$ evolve smoothly through the transition.

A further consistency check concerns the sign of $\dot H$. In a flat FRW universe within GR,
$\dot H=-\frac{\kappa c^2}{2}\,(\varepsilon_{\rm tot}+p_{\rm tot})
=-\frac{\kappa c^2}{2}\left(\varepsilon_{\rm m}+\frac{4}{3}\varepsilon_{\rm r}-2X\right)$,
so $\dot H>0$ would imply $\varepsilon_{\rm tot}+p_{\rm tot}<0$, i.e. violation of the null energy condition by the total cosmic medium and hence a super-accelerating phase~\cite{Caldwell:2025inn}. Because the phantom sector may be exotic while the full cosmic medium remains controlled, we require $\dot H<0$ over the cosmological history through the present. In the mirror phantom case the negative kinetic contribution is largest during the transition, so only this stage can threaten super-acceleration over that interval. A fast-transition estimate in Ref.~\cite{Akarsu:2025gwi}, obtained by linearizing the potential around $z_{\rm t}$, shows that imposing $\dot H(z\sim z_{\rm t})<0$ yields $z_{\rm t}\gtrsim 2(\Omega_{\rm m0}^{-1}-1)^{1/3}-1$. For $\Omega_{\rm m0}\sim0.27$, this gives $z_{\rm t}\gtrsim1.8$. In the solutions of Ref.~\cite{Akarsu:2025gwi}, the SH0ES-compatible case $z_{\rm t}\simeq2.1$ lies safely above this GR super-acceleration boundary, while approaching $z_{\rm t}\simeq1.8$ drives $\dot H(z_{\rm t})$ toward zero and makes the inferred $H_0$ highly sensitive to the transition epoch. Thus the requirement $\dot H<0$ through the present defines not only a theoretical lower bound within this GR realization but also helps explain why viable $H_0$-enhancing solutions occupy a narrow transition range.

\begin{figure}[!t]
    \centering
    \includegraphics[width=0.47\linewidth]{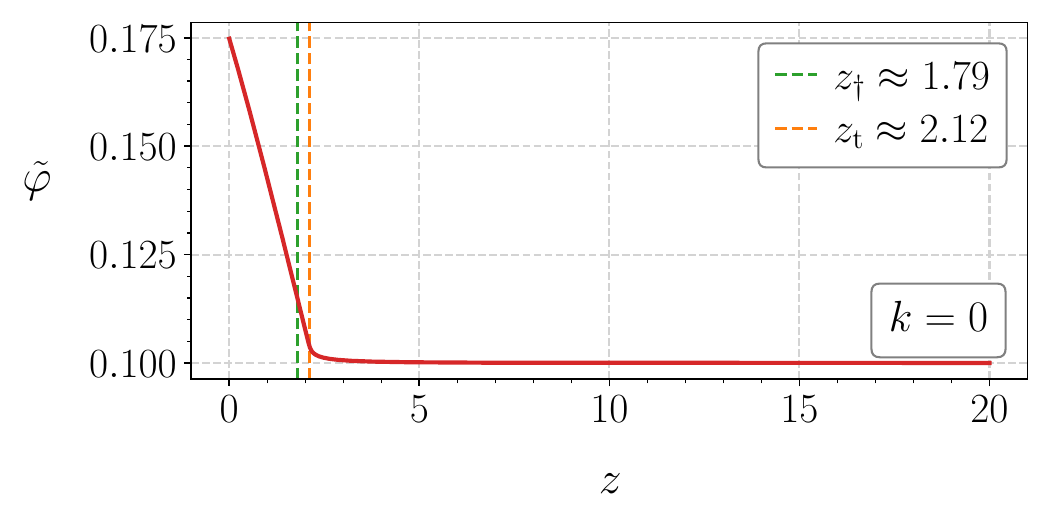}\hfill
    \includegraphics[width=0.47\linewidth]{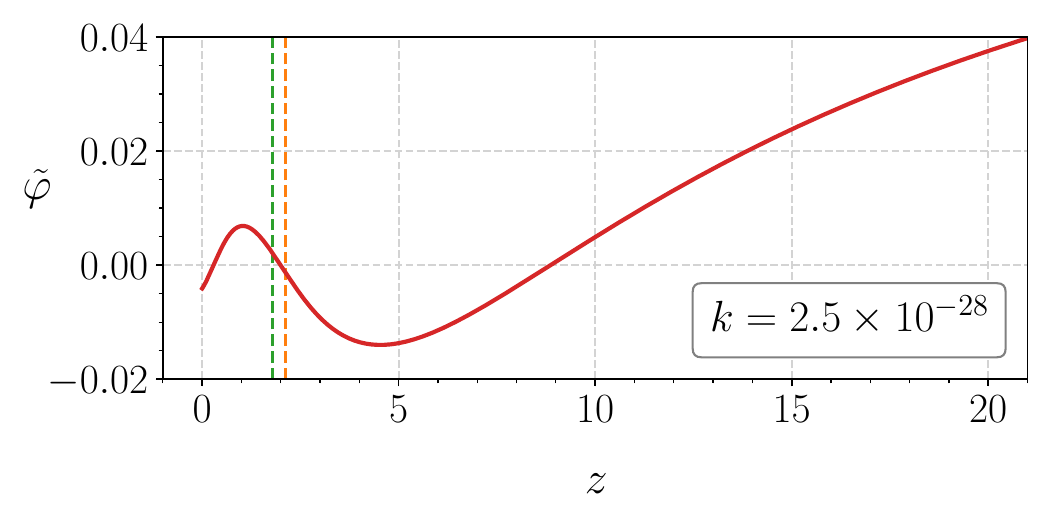}\\[0.75em]
    \includegraphics[width=0.47\linewidth]{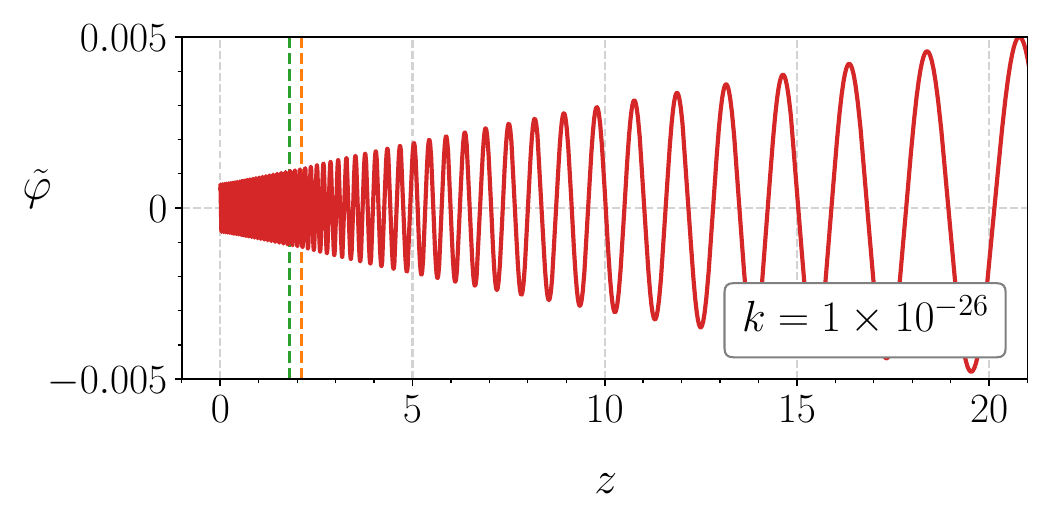}
    \caption{(Color online) Evolution of the dimensionless scalar-field perturbation $\tilde{\varphi}(z;k)=\tilde{\phi}(z;k)-\tilde{\phi}_{\rm bg}(z)$ for the mirror-transition benchmark $\alpha=0$ and $\nu=100$. The top-left, top-right and bottom panels correspond, respectively, to $k=0$, $k=2.5\times10^{-28}~{\rm cm^{-1}}$ and $k=1.0\times10^{-26}~{\rm cm^{-1}}$. The first two modes lie below $k_{\rm max}$ near the transition, whereas the last lies above it. The vertical dashed lines mark the density sign-switch redshift $z_{\dagger}\approx1.79$ (green) and the transition redshift $z_{\rm t}\approx2.12$ (orange). Initial conditions are $\tilde{\varphi}_{\rm in}=0.1$ and $\tilde{\varphi}_{\rm in}'=0$.}
    \label{fig:pert_nu100}
\end{figure}

The only dynamically delicate stage is the transition itself. At the perturbative level, linear phantom fluctuations in Fourier space are characterized by an effective mass squared
\begin{equation}
\label{eq:meff}
m^2_{\rm eff}(z)=\frac{k^2}{a^2}-\left.\frac{{\rm d}^2 V}{{\rm d}\phi^2}\right|_{\phi_{\rm bg}}
= k^2(1+z)^2-\frac{\varepsilon_{\rm cr0}}{M^2_{\rm Pl}}
\left.\frac{{\rm d}^2 \tilde\Omega_V}{{\rm d}\tilde\phi^2}\right|_{\tilde\phi_{\rm bg}}\, ,
\end{equation}
where $\phi_{\rm bg}(z)$ is the background field and we introduced the dimensionless functions $\tilde\Omega_V=V/\varepsilon_{\rm cr0}$ and $\tilde\phi=\phi/M_{\rm Pl}$. Because the kinetic term carries the opposite sign from the canonical case, a region with positive potential curvature contributes tachyonically. Hence, whenever $\mathrm{d}^2V/\mathrm{d}\phi^2>0$, a transient tachyonic band, $m_{\rm eff}^2<0$, opens for sufficiently small comoving wave numbers, $k^2<\frac{\varepsilon_{\rm cr0}}{M^2_{\rm Pl}(1+z)^2}
\left.\frac{{\rm d}^2 \tilde\Omega_V}{{\rm d}\tilde\phi^2}\right|_{\tilde\phi_{\rm bg}}$. For the mirror-transition case $\alpha=0$ and $\nu=100$, the appreciable positive-curvature region relevant to the transient tachyonic band lies approximately in the narrow range $z_{\rm t}<z\lesssim2.3$~\cite{Akarsu:2025gwi}. A conservative upper envelope for the instantaneous comoving-wave number threshold, obtained from the peak positive curvature, is
\begin{align}
\label{eq:kmax}
k_{\rm max}(z)
=\frac{2H_0\nu}{c(1+z)}\sqrt{\frac{\Omega_{\Lambda0}}{\sqrt{3}}}
\approx \frac{\nu H_0}{c(1+z)}\,1.30
\approx \frac{\nu}{1+z}\,1.03\times 10^{-28}\,{\rm cm}^{-1}\, ,
\end{align}
where we used $H_0=73.04~{\rm km\,s^{-1}\,Mpc^{-1}}$ and $\Omega_{\Lambda0}=0.7316$. Using $1+z\simeq1+z_{\rm t}=3.12$ across this narrow interval, Eq.~\eqref{eq:kmax} gives the overall bound $k\lesssim3.3\times10^{-27}\,{\rm cm}^{-1}$, or $\lambda_{\rm com}=2\pi/k\gtrsim6.2\times10^2\,{\rm Mpc}$. Thus any unstable modes are confined to extremely large wavelengths. They are also short-lived: $\Delta z\approx0.2
\quad\Longrightarrow\quad
\Delta t\sim \frac{\Delta z}{(1+z)H(z\sim2.2)}
\simeq2.2\times10^{-2}H_0^{-1}$. 
Here radiation is negligible, so
$H(z)/H_0\simeq[\Omega_{\rm m0}(1+z)^3+
\widetilde\Omega_\phi(z)]^{1/2}$. With
$\Omega_{\rm m0}\simeq0.2683$ and the pre-transition approximation
$\widetilde\Omega_\phi\simeq-\Omega_{\Lambda0}\simeq-0.7316$, this gives
$H(z\simeq2.2)\simeq2.84H_0$~\cite{Akarsu:2025gwi}.
The potentially unstable modes are therefore infrared and active only for a small fraction of a Hubble time. Figure~\ref{fig:pert_nu100} confirms that, for the modes and initial conditions shown, this transient tachyonic window produces only mild amplification. The perturbations remain smooth and continuous throughout the evolution, including at the redshift $z_{\dagger}$ where $\varepsilon_\phi=0$. Even in the most conservative case $k=0$, an initial perturbation amplitude $\tilde{\varphi}_{\rm in}=0.1$ grows only to $\max\tilde{\varphi}\simeq0.175$. Because the linearized perturbation equation is homogeneous, the choice $\tilde{\varphi}_{\rm in}=0.1$ fixes only the plotting normalization; the physical result is the modest amplification factor $\max|\tilde{\varphi}|/|\tilde{\varphi}_{\rm in}|\simeq1.75$. Any initial fluctuation small enough for the linearization therefore remains perturbative across the transition for the modes examined.

A separate and more familiar objection to phantom theories concerns quantum vacuum instability. Even if direct couplings to Standard Model fields are suppressed, graviton-mediated processes can in principle trigger vacuum decay~\cite{Carroll:2003st,Cline:2003gs}. The natural interpretation of such a sector is therefore that of an effective theory valid below a cutoff $\Lambda_{\rm cut}$. The background scales must lie below $\Lambda_{\rm cut}$, while vacuum decay bounds it from above: the optimistic estimate of Ref.~\cite{Carroll:2003st} permits $\Lambda_{\rm cut}\lesssim100~{\rm MeV}$, whereas the diffuse gamma-ray background gives $\Lambda_{\rm cut}\lesssim3~{\rm MeV}$ in the Lorentz-violating-cutoff treatment of Ref.~\cite{Cline:2003gs}. This low upper bound is a serious limitation, not evidence for a satisfactory ultraviolet completion.

Our model operates vastly below even the most conservative of these scales. First, since $\Lambda=\Omega_{\Lambda0}\,\varepsilon_{\rm cr0}$ and $\Lambda/M_{\rm Pl}^4\simeq9.0\times10^{-121}$ for the reduced Planck mass used here, it follows that $\Lambda^{1/4}\sim10^{-3}\,{\rm eV}$. Thus, in both the AdS-like and dS-like plateaus, where the potential is flat and cosmic friction nearly freezes the field, the dynamics are already deep in the infrared. The only potentially sensitive regime is the transition itself. The peak curvature scale of the potential, attained near $z\approx z_{\rm t}$, is
\begin{equation}
\label{eq:secondderivative}
\left|\frac{{\rm d}^2 V}{{\rm d} \phi^2}\right|_{\rm max}^{1/2}
= \sqrt{\frac{\varepsilon_{\rm cr0}}{M^2_{\rm Pl}}}
\left|\frac{{\rm d}^2 \tilde\Omega_V}{{\rm d} \tilde\phi^2}\right|_{\rm max}^{1/2}
=\nu\left[\frac{2|2-\alpha|}{3\sqrt{3}}\right]^{1/2}\frac{\Lambda^{1/2}}{M_{\rm Pl}}\, .
\end{equation}
For the mirror-transition case $\alpha=0$ and $\nu=100$, Eq.~\eqref{eq:secondderivative} gives a peak curvature scale of $2.0\times10^{-31}\ {\rm eV}\approx130\,H_0$ in natural units (using $\hbar H_0\simeq1.56\times10^{-33}\ {\rm eV}$), consistent with the peak-curvature envelope in Eq.~\eqref{eq:kmax}. Numerical integration gives $\max|{\rm d}\tilde\phi/{\rm d}z|\sim0.3$ at $z\simeq2.05$, shortly after the potential inflection point at $z_{\rm t}\simeq2.12$~\cite{Akarsu:2025gwi}. Returning to dimensional units, $|{\rm d}\phi/{\rm d}t| = M_{\rm Pl}|{\rm d}\tilde\phi/{\rm d}z|(1+z)H(z)$. Using $H(z)$ at this same redshift gives the characteristic peak scale $|{\rm d}\phi/{\rm d}t|_{\rm max}\sim1.0\times10^{-5}\ {\rm eV}^2$.
Both the characteristic curvature scale and the dynamical field-velocity scale are therefore many orders of magnitude below $\Lambda_{\rm cut}$ and $\Lambda_{\rm cut}^2$, respectively. Together, the estimates give $\Lambda^{1/4}$, $|V_{,\phi\phi}|_{\rm max}^{1/2}$, and $|\dot\phi|_{\rm max}^{1/2}\ll\Lambda_{\rm cut}\lesssim3~{\rm MeV}$. This separation supports the model as a classical low-energy proof of concept but does not remove the vacuum-instability or ultraviolet-completion caveats associated with the wrong-sign kinetic term.\footnote{At the limited phenomenological level of a lifetime constraint, the situation resembles proton-decay constraints in grand unified theories: a predicted instability need not exclude an effective description provided that its rate satisfies the relevant observational bounds. The analogy does not extend to the microscopic mechanism or the ultraviolet-completion problem.}

Having separated the effective-theory caveat from the background dynamics, we can now state precisely what the transition denotes. In this work, ``AdS-to-dS'' refers to the sign change of the DE contribution, not to a transition between two global spacetime vacua: the matter- and radiation-dominated FRW background is not assumed to tunnel from AdS to dS. The transition is instead an entirely classical solution of the coupled homogeneous Einstein--Klein--Gordon system, Eqs.~\eqref{eq:kg} and \eqref{eq:friedmann-raychaudhuri}. On the negative plateau, both $V_{,\phi}$ and the local curvature $V_{,\phi\phi}$ are exponentially suppressed, while the term $3H\dot\phi$ strongly damps the motion in the early universe. The field therefore remains nearly frozen and mimics a negative cosmological constant. As the expansion rate decreases, the phantom slowly climbs toward the step; once it reaches the shoulder, the reversed force drives the rapid but smooth passage to the positive plateau. The epoch of this delayed evolution is controlled by the initial displacement from the step (equivalently, the step location relative to the initial field value), whereas $\nu$ primarily controls its rapidity. For the benchmark $\nu=100$ solution of Ref.~\cite{Akarsu:2025gwi}, this displacement was calibrated by imposing the CMB-distance and physical-matter-density anchors together with the adopted local $H_0$ value. In the shifted coordinate in which the step is at $\phi=0$, the calibrated initial conditions are $\tilde\phi_{\rm in}\equiv\phi_{\rm in}/M_{\rm Pl}\simeq-0.0342$ and $\left.\mathrm{d}\tilde\phi/\mathrm{d}z\right|_{\rm in}=0$ at $z_{\rm in}=z_*\simeq1090$; they yield $z_{\rm t}\approx2.12$. Thus the transition epoch is not a parameter-free prediction of GR. The background sign change is not produced by tunneling or vacuum decay and requires no quantum trigger.

We can now state the main result succinctly. Ph-$\Lambda_{\rm s}$CDM realizes a smooth mirror AdS-to-dS transition within GR. By suppressing $H(z)$ before the sign switch and enhancing it afterward, the model reproduces the basic $\Lambda_{\rm s}$CDM mechanism for raising $H_0$ while preserving the CMB distance anchor. Within the background and linear analysis presented here, the full cosmic medium keeps $\varepsilon_{\rm tot}>0$ and $\omega_{\rm tot}>-1$ through the present, the spacetime approaches de Sitter asymptotically rather than encountering a Big Rip, the tachyonic modes examined acquire only the modest growth factor shown in Fig.~\ref{fig:pert_nu100}, and the characteristic background scales remain deep in the infrared relative to the quoted effective-theory cutoffs.

For this reason, Ph-$\Lambda_{\rm s}$CDM~\cite{Akarsu:2025gwi} is more than an exotic construction. It supplies a concrete GR underpinning for the broader $\Lambda_{\rm s}$CDM~\cite{Akarsu:2019hmw,Akarsu:2021fol,Akarsu:2022typ,Akarsu:2023mfb} idea and sharpens the very meaning of phantom and repulsive DE: once the density is allowed to change sign, the phantom-divide line ceases to be a global separator, and repulsion need not await positive DE density. Phenomenologically, the same mechanism may help alleviate not only the Hubble tension but the wider pattern of late-time tensions; its emerging connection to the growth-index $\gamma$ tension, together with the enhanced structure growth implied by negative pre-transition DE~\cite{Paraskevas:2024ytz,Akarsu:2025ijk,Escamilla:2025imi,Akarsu:2025nns}, also suggests a natural link to JWST-era high-redshift anomalies, including the so-called impossibly early galaxy problem~\cite{Steinhardt:2015lqa,Boylan-Kolchin:2022kae,Matthee:2023utn,Rusakov2026}. Even if some current tensions ultimately prove to be systematic, their persistence in the precision-cosmology era makes it timely to explore minimal alternatives to $\Lambda$CDM that connect microphysical dynamics to cosmological phenomenology. But the central gravitation lesson is simpler and deeper: within GR, a bounded phantom sector can realize a controlled sign reversal of DE from an AdS-like past to a dS-like present while maintaining a regular, nonsingular FRW background.
In this light, the phantom ceases to be a \emph{menace} to the stability of the cosmos; instead, it emerges as a friendly component of a consistent cosmological framework.

%\vspace{0.5cm}

\begin{acknowledgments}
The authors acknowledge networking support and participation in the CosmoVerse network.
\end{acknowledgments}

\begin{samepage}
\section*{ORCID}
\noindent
\"{O}zg\"{u}r Akarsu\,\orcidlink{0000-0001-6917-6176}\quad\url{https://orcid.org/0000-0001-6917-6176}\\
Leandros Perivolaropoulos\,\orcidlink{0000-0001-9330-2371}\quad\url{https://orcid.org/0000-0001-9330-2371}\\
A. Emrah Y\"{u}kselci\,\orcidlink{0000-0003-4554-1268}\quad\url{https://orcid.org/0000-0003-4554-1268}\\
Alexander Zhuk\,\orcidlink{0000-0001-9107-5048}\quad\url{https://orcid.org/0000-0001-9107-5048}
\par
\end{samepage}

\end{document}